\documentclass[aps, prr, reprint, superscriptaddress]{revtex4-2}
\usepackage{graphicx}  
\usepackage{dcolumn}   
\usepackage{bm}        
\usepackage{amssymb}   
\usepackage{color}
\usepackage{amsmath}
\usepackage{amsfonts, amsmath, amssymb}
\usepackage{appendix}
\usepackage[dvipsnames]{xcolor}
\usepackage{tikz}
\usepackage{graphicx}
\begin{document}

\title{Approximate Analytical Protein Distributions for the Three-stage Model of Stochastic Gene Expression}

\author{Kenny Wong}
\email{\texttt{wong.1216@osu.edu}}
\affiliation{Department of Physics, The Ohio State University, Columbus, Ohio 43210}

\author{Thomas Mourier}
\affiliation{Department of Mechanical and Industrial Engineering, Northeastern University, Boston, MA 02115}

\author{Sho Inaba}
\affiliation{Department of Physics, University of Massachusetts Boston, Boston, MA 02125}

\author{Cameron Hopkinson}
\affiliation{Department of Biology, Washington University, St. Louis, MO 63130}

\author{Rahul Kulkarni}
\email{\texttt{rahul.kulkarni@umb.edu}}
\affiliation{Department of Physics, University of Massachusetts Boston, Boston, MA 02125}

\begin{abstract}
    Gene expression is an intrinsically stochastic process that generates phenotypic heterogeneity within genetically identical cell populations. While the exact statistical moments of the protein count can be obtained for a broad range of complex models, the corresponding distributions are significantly harder to obtain and intractable in many cases. The classical three-stage model of gene expression, which predicts fluctuations in protein levels as a function of promoter switching, transcription, translation, and degradation events all occurring with linear propensities, illustrates this perfectly; deriving its exact protein distribution remains elusive. Here, {using the partitioning property of time-inhomogeneous Poisson processes, we develop an exact mapping of the three-stage model onto a simplified model.} The simplified model allows us to formulate two analytical approximations for the full protein distribution of the three-stage model based on a beta-mixture representation of the exact solution for a simpler model. We show that the two approximations are asymptotically exact in different limiting cases and verify their accuracy against simulations for a broad range of parameters. Although approximate, these are the first analytical expressions for protein distributions for the three-stage model that are highly accurate in intermediate regimes. 
\end{abstract}

\maketitle

\section{Introduction}
Modern biological research has overwhelmingly shown that gene expression is an intrinsically random process and numerous experiments have reported a remarkable level of cell-to-cell variability in both mRNA and protein counts in genetically identical cell populations \cite{Raj-Cell-2008}. Stochasticity (noise) in gene expression creates nongenetic phenotypic heterogeneity which underpins many important biological phenomena involving cell fate decisions, ranging from drug tolerance in melanoma \cite{Shaffer-Nature-2017, Schuh-CellSys-2020} to bacterial persistence \cite{Mirouze-plos-2011}. Thus, there is great interest in modeling the underlying processes of stochastic gene expression to quantify noise and investigate the properties of the corresponding mRNA and protein distributions to aid in our understanding of phenotypic variation in genetically identical cells. 

Gene expression is commonly analyzed through coarse-grained stochastic models such as the two and three-stage models \cite{Paulsson-PLR-2005} depicted in the top panels of Figures 1C and 2 respectively. The two-stage model assumes a constitutively active promoter and represents translation and degradation as first order reactions. The three-stage model extends this to allow for the promoter to toggle between an active and inactive state. Although these models give a simplified depiction of the processes underlying gene expression, they serve as building blocks for more complex models that incorporate complex promoter-based \cite{Sanchez-PNAS-2008, Zhou-SIAM-2012}, post-transcriptional \cite{Jia-PRL-2010, Wong-PRE-2026, Wong-PhysBiol-2026}, or feedback regulatory mechanisms \cite{Hornos-PRE-2005, Kumar-PRL-2014}. The strength of cellular fluctuations in the mRNA and protein count can be characterized by the mean and variance of their corresponding distributions. While the exact mean and variance for the broad class of gene expression models that are composed of entirely zeroth and first order reactions (including the two and three-stage models) can be easily obtained (because the moment equations are always closed), the corresponding analytic solutions for the full distributions {for protein counts} are extremely difficult to find and thus lacking in many cases. Many protein distributions exhibit features such as bimodality which are not adequately characterized by their first two moments alone. Given that protein levels can be accurately measured in single cells \cite{Vistain-biochemsci-2021}, analytical distributions can produce testable predictions for experiments and are particularly useful for inverse problems such as parameter inference. 

Notably, it took over a decade of extensive theoretical work to obtain the exact analytical solution for the steady-state protein distribution of the two-stage model \cite{Bokes-JMathBio-2012} and the corresponding solution for the three-stage model remains an outstanding open problem. {Using the partitioning property of Poisson arrivals (PPA), we have previously developed an analytical mapping in prior work which enables exact protein distributions to be directly obtained by leveraging mRNA distributions \cite{Pendar-PRE-2013} corresponding to simplified systems. However, the PPA mapping has only been applied on models where transcription follows a Poisson process (i.e. constitutively active promoter). Given that the partitioning property still holds for \textit{time-inhomogeneous} arrivals, we develop an extension of the PPA mapping for the three-stage model.} This allows us to analytically derive two approximate distributions which take on a beta-mixture functional form that become asymptotically exact in different limiting cases. Although approximate, to our knowledge, we give the first analytical expressions for the steady-state protein distribution of the three-stage model that show excellent agreement with simulations in intermediate parameter regimes without needing to restrict any part of the dynamics to a limiting timescale.

\section{Mapping to Reduced Models}
\begin{figure}
    \centering
    \includegraphics[width=1\linewidth]{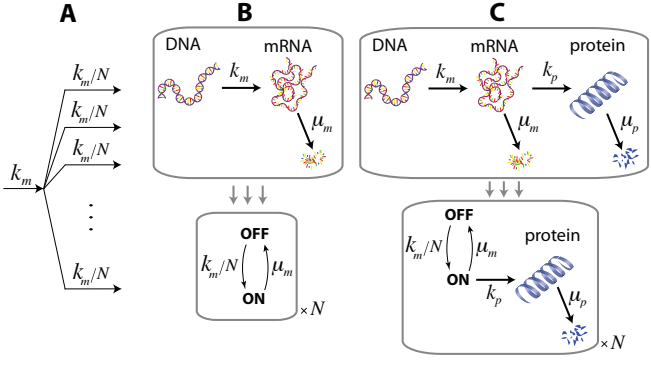}
    \caption{Adapted from Ref. \cite{Pendar-PRE-2013}. A) Poisson arrival process with arrival rate $k_m$ is partitioned into $N$ independent and identical Poisson arrival processes, each occurring with rate $k_m/N$. B) Partitioning of the Poisson arrival process leads to a mapping from a simple model of creation and decay of mRNAs to $N$ independent, identical two-state systems in the limit $N \to \infty$. The probability of having $m$ mRNAs in the original model is equivalent to the probability of having $m$ two-state systems in the ON state in the reduced model. C) The same mapping applied to the two-stage model of gene expression for proteins. The reduced model is identical to a model for creation and decay of mRNAs with promoter-based regulation.}
\end{figure}
Let $P(S,t)$ represent the probability of the system being in state $S$ at time $t$. In the context of gene expression models, $S$ is typically characterized by a set of integers $\{s_1, s_2, \dots, s_j\}$ which represent the random variables of interest such as the promoter state, mRNA, or protein count. $P(S,t)$ evolves according to the master equation
\begin{equation}
    \frac{\partial P(S,t)}{\partial t} = \sum_{S'} \big\{ W_{S' \rightarrow S} P(S',t) - W_{S \rightarrow S'} P(S,t) \big\}
\end{equation}
where $W_{S' \rightarrow S}$ represents the rate at which the system transitions from state $S'$ to state $S$. The master equation(s) are usually solved in their generating function form, which are obtained through the transformation 
\begin{align}
    G(\{&z_1, z_2, \dots, z_j\},t) =  \notag \\
    &\sum_{\{s_1, s_2, \dots, s_j\}} z_1^{s_1} z_2^{s_2}\dots z_j^{s_j} P(\{s_1, s_2, \dots, s_j\},t).
\end{align}
The probability distribution $P(S,t)$ is encoded in the coefficients of the Taylor series of $G$ and the corresponding moments of the distribution can be found by successive differentiation. {We review how $G$ can be obtained using the PPA mapping on the two-stage model for both the mRNA and protein level following along Ref. \cite{Pendar-PRE-2013}.}

{Let us begin by partitioning the mRNA arrivals into $N$ types (Figure 1A) with each unique type corresponding to an individual partition. Each mRNA arrival is randomly assigned to an arbitrary type $i$ drawn from $\{ 1, 2, \dots, N \}$ with uniform probability $q_i = 1/N$. From the partitioning theorem of Poisson processes, the arrival processes of mRNAs for each type are each independent, identical Poisson process with a rescaled arrival rate $k_m/N$. Correspondingly, the number of arrivals up to time $t$ in each partition are independent, identically distributed random variables drawn from Poisson$(k_mt/N)$. Each partition independently contributes to the total number of mRNA arrivals $M = \sum^N_{i=1} m_i$ where $m_i$ corresponds to number of mRNA arrivals of the $i$-th type. This logic also applies to any random variable of interest. Thus, at any time $t$, we can map the dynamics of the original model onto the dynamics of $N$ identical subsystems.}

{Now, in the limit $N \to \infty$, the probability of each partition having more than one mRNA arrival can be neglected (because it is infinitely more likely that the additional arrivals get assigned to any other type). This constrains the random variable $m$ corresponding to the mRNA count in each partition to be either 0 or 1, replacing the original Poisson mRNA dynamics with a simple two-state system (Figure 1B) which we refer to as the \textit{reduced} model. Recalling that the dynamics of each partition are independent of one another, the generating function of the reduced model $g(z,t)$ can be used to construct that of the original model by
\begin{equation}
    G(z,t) = \lim_{N \to \infty} [g(z,t)]^N.
\end{equation}
Working in the limit $N \to \infty$ only requires expressions to be in terms of leading order $1/N$ for exactness. Consequently, we have previously shown that $g(z,t)$ can be alternatively related to $G(z,t)$ by 
\begin{equation}
    G(z,t) = \lim_{N \to \infty} \exp \{ N(g(z,t)-1) \}.
\end{equation}
Returning to the reduced model under consideration (Figure 1B), the dynamics are defined by transitions between an OFF ($m=0$) and ON state ($m=1$) at rates $k_m/N$ and $\mu_m$ respectively. Thus, the stationary generating function is linear in terms of $z$
\begin{equation}
    g(z) = \frac{\mu_m}{\mu_m+k_m/N} + z\frac{k_m/N}{\mu_m+k_m/N}.
\end{equation}
Substituting this into Eq. (4), it is easy to see that $G(z)$ becomes the stationary Poisson generating function with mean $k_m/\mu_m$, recovering a trivial result.}

{The preceding argument holds for protein distributions too. We can likewise construct a reduced model for the protein dynamics for the two-stage model depicted in Figure 1C, which resembles transcription undergoing promoter-based regulation, referred to as the telegraph model. The stationary mRNA generating function for the telegraph model has been derived in prior work \cite{Raj-plos-2006}
\begin{equation}
    g(z) = {_1}F_1 \bigg[ \frac{k_m/N}{\mu_p}; \frac{\mu_m}{\mu_p}; \frac{k_p}{\mu_p}(z-1) \bigg]
\end{equation}
where ${_1}F_1$ represents the confluent hypergeometric function. This leads to a very quick derivation of the protein distribution for the two-stage model
\begin{align}
    &G_{\text{two-stage}}(z) = \\
    &\lim_{N\rightarrow \infty} \exp \bigg\{ N \bigg( {_1F_1} \bigg[\frac{k_m/N}{\mu_p}; \frac{\mu_m}{\mu_p}; \frac{k_p}{\mu_p}(z-1)  \bigg] - 1 \bigg) \bigg\} \notag .
\end{align}
The power of the PPA mapping lies in the fact that exact protein distributions can be readily obtained by leveraging results for mRNA distributions (pre-existing in many cases) since the reduced models are mathematically identical to models of promoter-based regulation. In the following section, we adapt the PPA mapping for time-inhomogeneous Poisson mRNA arrivals in the three-stage model due to promoter switching.}

\section{Time-inhomogeneous PPA Mapping for the Three-stage Model}
{Consider the reduced model for the three-stage model depicted in Figure 2. Since the partitioning property still holds for time-inhomogeneous Poisson processes, the dynamics of each partition are likewise identical and independent conditioned on the same trajectory of $\xi(t)$ so the arguments made in the last section about the mapping between the reduced model and the original model also hold in this setting. Here, the mRNA arrival rate now varies as a function of time $\xi(t) k_m/N$ where $\xi(t) \in \{0,1\}$ is a dichotomous random variable which indicates the underlying promoter state.}

\begin{figure}
    \centering
    \includegraphics[width=0.9\linewidth]{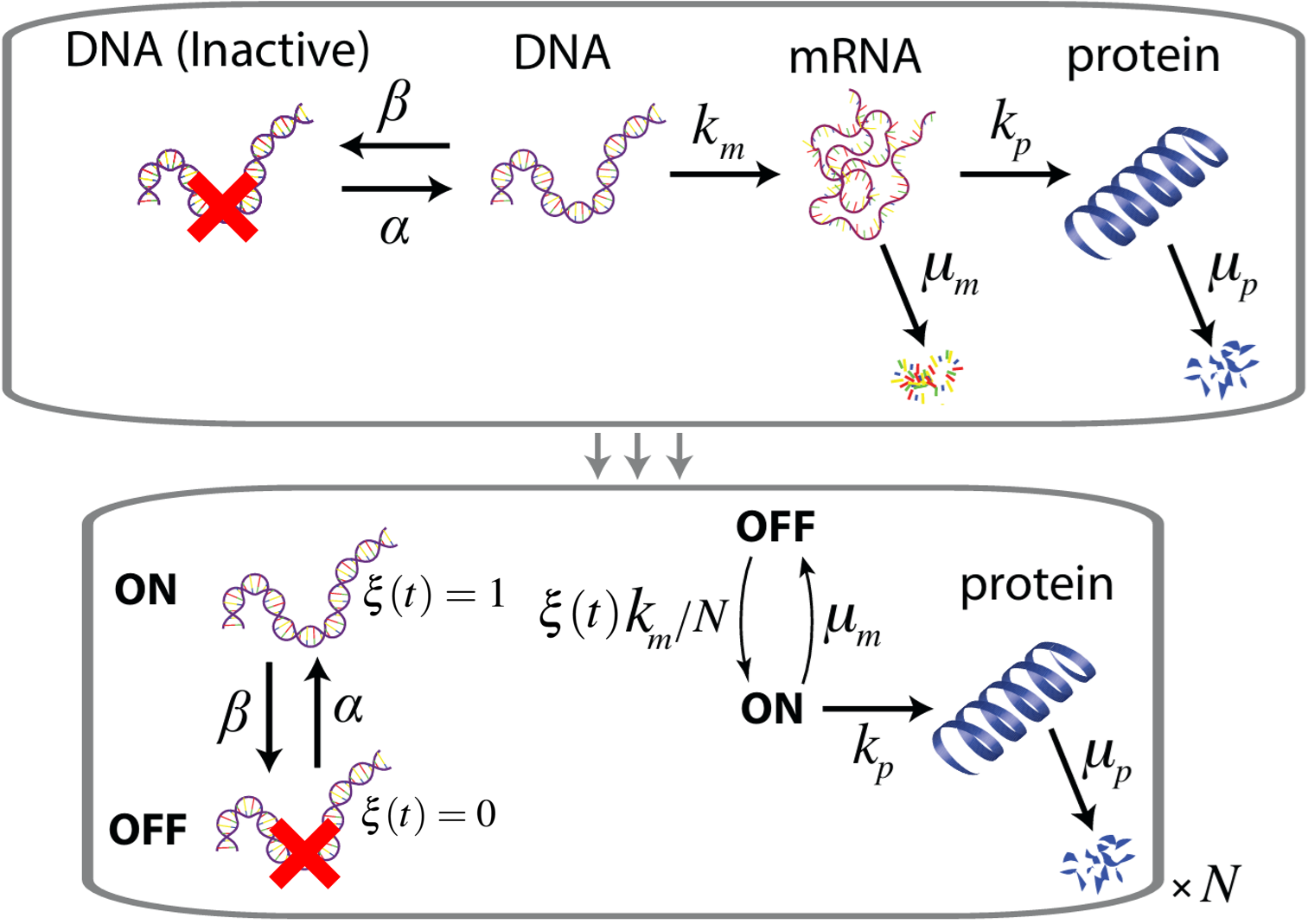}
    \caption{The PPA mapping adapted for the three-stage model. The reduced model is nearly identical to that of the two-stage model with the mRNA arrival rate being $\xi(t) k_m/N$ where $\xi(t)$ indicates the underlying promoter state $\xi(t)=0$ for OFF and $\xi(t)=1$ for ON. Since the partitioning property also holds for time-inhomogeneous Poisson arrivals, the dynamics of the reduced models are likewise identical and independent conditioned on a fixed realization of $\xi(t)$.}
    \label{figplaceholder}
\end{figure}

\subsection{mRNA Distribution}
Let us first illustrate the time-inhomogeneous PPA mapping by using it to recover the exact mRNA distribution of the three-stage model. Once again, the key simplification is that we need only to consider the dynamics of one mRNA. \textit{Conditioned} on a fixed realization of the trajectory of $\xi(t)$ up to time $t$, the generating function can be constructed as
\begin{align}
    g_{\xi} (z,t) = \big[ 1 - M(1,t) \big] + zM(1,t)
\end{align}
with $M(1,t)$ being the probability that the mRNA has arrived before time $t$ but has not yet degraded
\begin{align}
    M(1,t) &= \frac{k_m}{N} \int_0^t ds\, \xi(s) e^{-\frac{k_m}{N}\int_0^s  ds'\, \xi(s') } e^{-\mu_m(t-s)}  \notag \\
    &= \frac{k_m}{N} \int_0^t ds\, \xi(s) e^{-\mu_m(t-s)} 
\end{align}
where the simplification is from dropping terms beyond leading order $1/N$. The generating function for the original model conditioned on the same trajectory of $\xi(t)$ is
\begin{equation}
    G_\xi(z,t) = \exp\bigg\{ \frac{k_m}{\mu_m}\lambda_t (z-1) \bigg\}
\end{equation}
where we have defined
\begin{equation}
    \lambda_t = \int_0^t ds \, \xi(s) \mu_m e^{-\mu_m(t-s)} .
\end{equation}
Notably, Eq. (10) is the generating function of a Poisson process of mean $\frac{k_m}{\mu_m}\lambda_t$ with $0 \leq \lambda_t \leq 1$. $\lambda_t$ can be seen as the weighted average of the promoter history according to an exponential distribution. To obtain the exact mRNA distribution for the original model, we must average over all possible realizations of the promoter history encoded in $\lambda_t$ 
\begin{equation}
    G(z,t) = \int_0^1 d\lambda_t \, p(\lambda_t,t) \exp\bigg\{ \frac{k_m}{\mu_m}\lambda_t (z-1) \bigg\} , 
\end{equation} 
which we will carry out in the stationary limit, with $p(\lambda_t,t)$ being the distribution of $\lambda_t$. This representation of $G$ indicates that the mRNA distribution can be seen as a continuous superposition of a Poisson process in which the Poisson parameter itself (through $\lambda_t$) is stochastic; this is referred to as a Cox process.  

To obtain the distribution for the Poisson parameter, let us define the CDFs
\begin{align}
    P_0(\lambda, t) = \text{Prob}[\lambda_t \leq \lambda, \xi(t) = 0] \\
    P_1(\lambda, t) = \text{Prob}[\lambda_t \leq \lambda, \xi(t) = 1]
\end{align}
where $\lambda$ denotes a realization of the random variable $\lambda_t$ and the CDFs, over an infinitesimal timestep, evolve according to
\begin{align}
    P_0(\lambda, t+dt) &= P_1(\lambda, t) \beta \, dt + P_0(\tilde{\lambda}, t) \big[1-\alpha \, dt\big] \\ 
    P_1(\lambda, t+dt) &= P_0(\lambda, t) \alpha \, dt + P_1(\lambda', t) \big[1-\beta \, dt\big].
\end{align}
Now, $\tilde{\lambda}$ and $\lambda'$ at time $t$ are related to $\lambda$ at $t+dt$ by
\begin{align}
    \lambda = 
    \begin{cases}
        \tilde{\lambda} + d\lambda; \,\, &\xi(t) = 0 \\
        \lambda' + d\lambda; \,\, &\xi(t) = 1
    \end{cases}
\end{align}
in Eqs. (15) and (16) respectively and separately. Applying Leibniz's rule on the definition of $\lambda_t$, it is straightforward to show that
\begin{equation}
    \frac{d\lambda_t}{dt} = \mu_m(\xi(t) -\lambda_t)
\end{equation}
which leads to
\begin{align}
    \tilde{\lambda} &= \lambda(1+\mu_m dt); \, \, \, \lambda' = \lambda + \mu_m(\lambda -1) dt.
\end{align}
Substituting the preceding relations into Eqs. (15) and (16) then differentiating with respect to $\lambda$, we arrive at the corresponding master equations for the PDFs
\begin{align}
    \frac{\partial p_0(\lambda,t)}{\partial t} &= \beta p_1(\lambda,t) - \alpha p_0(\lambda,t) + \mu_m \frac{\partial}{\partial \lambda}\big[\lambda p_0(\lambda,t)\big]; \\
    \frac{\partial p_1(\lambda,t)}{\partial t} &= \alpha p_0(\lambda,t) - \beta p_1(\lambda,t) + \mu_m \frac{\partial}{\partial \lambda}\big[(\lambda-1) p_1(\lambda,t)\big] 
\end{align}
with $p_0(\lambda,t)+p_1(\lambda,t)=p(\lambda,t)$. After transforming the master equations into a single separable equation for the marginal distribution $p$ (Appendix A), we find that the stationary solution follows the beta distribution
\begin{equation}
    p(\lambda) = B\big(\lambda; a,b \big) = \frac{\Gamma(a+b)}{\Gamma(a)\Gamma(b)} (1-\lambda)^{b-1} \lambda^{a-1}
\end{equation}
where $\Gamma$ represents the gamma function, $a=\frac{\alpha}{\mu_m}$, $b=\frac{\beta}{\mu_m}$, and the prefactor was chosen to enforce probability conservation. Thus, the generating function of the stationary mRNA distribution can be exactly written as
\begin{equation}
    G(z) = \int_0^1 d\lambda \, B\bigg(\lambda; \frac{\alpha}{\mu_m}, \frac{\beta}{\mu_m}\bigg) \exp\bigg\{ \frac{k_m}{\mu_m}\lambda (z-1) \bigg\} 
\end{equation}
which is mathematically identical to \cite{NIST}
\begin{equation}
    G(z) = {_1F_1} \bigg[ \frac{\alpha}{\mu_m}; \frac{\alpha + \beta}{\mu_m}; \frac{k_m}{\mu_m}(z-1) \bigg],
\end{equation}
recovering a standard result. This beta-Poisson mixture representation of the stationary mRNA distribution for the telegraph model was originally noted in Ref. \cite{Jay-PRE-2009}. Here, we have derived this representation from first principles using the time-inhomogeneous PPA mapping. 
\\
\\

\subsection{Protein Distribution}
Following the same logic, the exact protein distribution of the three-stage model can be derived by first obtaining the protein generating function conditioned on a fixed trajectory of $\xi(t)$ and then likewise averaging over all possible realizations. The generating function for the reduced model, exact for leading order $1/N$, is
\begin{equation}
    g_\xi(z,t) = 1 + \frac{k_m}{N} \int_0^t ds\, \xi(s) \big[ f(z,s,t) -1 \big]
\end{equation}
where $f(z,s,t)$ is the protein generating function contribution for a single mRNA that arrived at $s < t$. From here, there are two cases to be considered which are independent of one another: (1) mRNA survives until $t$ with probability $e^{-\mu_m(t-s)}$; (2) mRNA degrades at $s' < t$ with probability $ds' \, \mu_m e^{-\mu_m(s'-s)}$. 

Let us define $X$ ($H_X(z,t)$), $Y$ ($H_Y(z,t)$), and $Z$ ($H_Z(z,t)$) to be the random variables (generating functions) corresponding to the number of proteins produced from case (1), from case (2), and between $s'$ and $t$ respectively such that $X=Y+Z$ and $H_X(z,t)=H_Y(z,t)H_Z(z,t)$. The contribution from case (1) corresponds to constitutively active translation so the protein counts trivially follow a Poisson distribution with a time-dependent mean
\begin{equation}
    H_X(z,t) = \exp\bigg\{ \frac{k_p}{\mu_p}(z-1) \big(1-e^{-\mu_p(t-s)}\big) \bigg\}.
\end{equation}
The same logic also applies for constructing $H_Z(z,t)$, so $H_Z$ is nearly identical to $H_X$ where $s$ is replaced with $s'$. Thus, the generating function for case (2) is
\begin{align}
    &H_Y(z,t) = \frac{H_X(z,t)}{H_Z(z,t)} \notag \\
    &= \exp\bigg\{ \frac{k_p}{\mu_p}(z-1)\big( e^{-\mu_p(t-s')} - e^{-\mu_p(t-s)} \big) \bigg\}.
\end{align}
Integrating over all possible $s'$ for case (2) and putting everything together, the exact generating function for the protein distribution of the three-stage model conditioned on a fixed trajectory of $\xi(t)$ is
\begin{equation}
    G_\xi(z,t) = \exp\bigg\{ \frac{k_m}{\mu_m} \int_0^t ds\, \xi(s) \mathcal{K}(z, t-s) \bigg\}
\end{equation}
where we define
\begin{widetext}
    \begin{align}
        \Lambda_t(z) := \int_0^t ds\, \xi(s) \mathcal{K}(z, t-s) = \int_0^t &ds\, \xi(s) \mu_m \Bigg[ \exp\bigg\{ \frac{k_p}{\mu_p}(z-1) \big( 1 - e^{-\mu_p(t-s)} \big) - \mu_m(t-s) \bigg\} \\ 
        &+\int_s^t ds'\, \mu_m \bigg( \exp\bigg\{ \frac{k_p}{\mu_p}(z-1)\big( e^{-\mu_p(t-s')} - e^{-\mu_p(t-s)} \big) -\mu_m(s'-s) \bigg\}\bigg) -1 \Bigg] \notag.
    \end{align}
    However, unlike the mRNA dynamics, it is easy to see that the expression for $d\Lambda_t(z)/dt$ to track the evolution of $\Lambda_t(z)$ does not admit a closed-form representation, making the corresponding master equations for the distribution of $\Lambda_t(z)$ (and thus the exact protein distribution of the three-stage model) analytically intractable via this approach. Instead, we seek to build an analytic \textit{approximation} with constraints from known exact results in limiting cases.
    
    To start, when $\xi(t)=1$ for all $t$ we must recover the exact steady-state protein distribution of the two-stage model, leading to the remarkable identity
    \begin{align}
          \lim_{t \rightarrow \infty} \frac{k_m}{\mu_m} \int_0^t ds\, \mathcal{K}(z, t-s) = \ln G_{\text{two-stage}}(z).
    \end{align}
    Given that $k_m$ only appears in one place on both sides of Eq. (30), if we scale $k_m \to k_m \lambda$, the relation 
    \begin{align}
          \lim_{t \rightarrow \infty} \frac{k_m \lambda}{\mu_m} \int_0^t ds\,  \mathcal{K}(z, t-s) = \lim_{N\rightarrow \infty} N \bigg( {_1F_1} \bigg[\frac{k_m \lambda/N}{\mu_p}; \frac{\mu_m}{\mu_p};\frac{k_p}{\mu_p}(z-1)  \bigg] - 1 \bigg)
    \end{align}
    holds. Similar to $\lambda_t$ defined in Eq. (11), $\Lambda_t(z)$ can be seen as a weighted sum of the promoter history but with the kernel $\mathcal{K}$ being a complicated, $z$-dependent weighting function instead of a simple exponential distribution. Let us define the ``normalized", stationary version of $\Lambda_t(z)$ as
    \begin{equation}
        \tilde{\Lambda}_z := \lim_{t\to\infty} \frac{\int_0^t ds\, \xi(s) \mathcal{K}(z, t-s)}{\int_0^t ds\, \mathcal{K}(z, t-s)}.
    \end{equation}
    Note that unlike $\lambda$, $\tilde{\Lambda}_z$ maps a trajectory of $\xi(t)$ onto a function of $z$ instead of a scalar. Now, the exact generating function for the stationary protein distribution can be expressed as the path integral
    \begin{align}
        G(z) = \int \mathcal{D} [\tilde{\Lambda}_z] \, p(\tilde{\Lambda}_z) \exp \big\{ \tilde{\Lambda}_z \ln G_\text{two-stage}(z) \big\}
    \end{align}
    where we integrate over all possible functions of $z$ pertaining to $\tilde{\Lambda}_z$ and $p$ is the distribution of $\tilde{\Lambda}_z$. Making use of the identities in Eq. (7) and then Eq. (31) allows us to further recast the generating function as
    \begin{align}
        G(z) = \lim_{N \to\infty} \int \mathcal{D}[\tilde{\Lambda}_z] \, p(\tilde{\Lambda}_z) \exp \bigg\{ N \bigg( {_1F_1} \bigg[ \frac{k_m \tilde{\Lambda}_z/N}{\mu_p} ; \frac{\mu_m}{\mu_p}; \frac{k_p}{\mu_p}(z-1)  \bigg] - 1 \bigg) \bigg\}.
    \end{align}
\end{widetext}
While this representation of the stationary protein distribution for the three-stage model is \textit{exact}, recall that the intractability lies in that $p(\tilde{\Lambda}_z)$ which does not have an analytic solution. To move forward, we must approximate $p(\tilde{\Lambda}_z)$ with a $z$-independent function.

Let us take $\tilde{\Lambda}_z \to \lambda$ which we define as
\begin{equation}
    \lambda = \lim_{t\to\infty} \int_0^t ds\, \xi(s) Ce^{-C(t-s)}  
\end{equation}
where the idea is to approximate the memory of the promoter history at the level of the protein dynamics with an exponential distribution. Since we have already shown $\lambda$ to follow the beta distribution, we formulate the approximate generating function ansatz
\begin{align}
    G&(z) = \lim_{N\rightarrow \infty} \int_0^1 d\lambda \, B\bigg(\lambda; \frac{\alpha}{C}, \frac{\beta}{C}\bigg)  \\
    &\times \exp \bigg\{ N \bigg( {_1F_1} \bigg[ \frac{k_m\lambda/N}{\mu_p} ; \frac{\mu_m}{\mu_p}; \frac{k_p}{\mu_p}(z-1)  \bigg] - 1 \bigg) \bigg\} \notag
\end{align}
where we set
\begin{equation}
    \frac{1}{C} = \frac{1}{\mu_m} + \frac{1}{\mu_p} - \frac{1}{\alpha + \beta + \mu_m + \mu_p}
\end{equation}
so that $G(z)$ reproduces the known mean and Fano factor for the three-stage model exactly \cite{Raser-Science-2004} (Appendix B1). Remarkably, $G(z)$ recovers the following limiting cases exactly: fast promoter switching ($\frac{\alpha}{\mu_p}; \frac{\beta}{\mu_p} \rightarrow \infty$); slow promoter switching ($\frac{\alpha+\beta}{\mu_p} \rightarrow 0$); slow mRNA dynamics ($\frac{\mu_m}{\mu_p} \rightarrow 0$). The asymptotic equivalences are shown in Appendix B2. $G(z)$ effectively approximates how the hidden layer of mRNA fluctuations driven by promoter switching propagate to the protein level and interpolates between multiple limiting parameter regimes.

\subsection{Fast mRNA Ansatz}
The shortcoming of the preceding approximation is that it breaks down in the limit $\frac{\mu_m}{\mu_p} \rightarrow \infty$ corresponding to fast mRNA dynamics. This is expected because the system dynamics under the fast mRNA limit are fundamentally different. Unlike the other asymptotic limits, the stochastic layer of the mRNA copy number, whose fluctuations are propagated to the protein level, is eliminated and replaced by a compound jump process where each mRNA arrival corresponds to an instantaneous burst of proteins drawn from a geometric distribution \cite{Shahrezaei-PNAS-2008} (with mean $\frac{k_p}{\mu_m}$). We refer to this as the translational burst approximation. Consequently, fluctuations in the mRNA copy number effectively disappear and get re-encoded as fluctuations in the protein burst sizes; we need to develop a separate generating function that recovers this limiting case. The stationary protein distribution here follows a Gaussian hypergeometric function ${_2F_1}$. To distinguish between the expressions, disconnected by the mRNA timescale, let us label the generating function which exactly recovers slow mRNA dynamics in Eq. (36) as $G_s(z)$ and the generating function to account for fast mRNA dynamics as $G_f(z)$. 

Interestingly, the Euler integral representation of the ${_2F_1}$ solution of the protein distribution in the fast mRNA limit in Ref. \cite{Shahrezaei-PNAS-2008} is a beta-mixture of the generating function of a negative binomial distribution, and the negative binomial generating function can be seen as a limiting case of ${_1F_1}$. With this in mind, we extrapolate the fast mRNA ansatz (working Appendix B3 backwards)
\begin{align}
    G&_f(z) = \lim_{N\rightarrow \infty} \int_0^1 d\lambda \, B\big(\lambda; a, b \big)  \\
    &\times \exp \bigg\{ N \bigg( {_1F_1} \bigg[\frac{k/N}{\mu_p}; \frac{\mu_m}{\mu_p}; \frac{k_p}{\mu_p}\lambda (z-1) \bigg] - 1 \bigg) \bigg\} \notag
\end{align}
where the key difference is that $\lambda$ now multiplies $k_p$ instead of $k_m$ and the set of parameters $\{ k$, $a$, $b \}$ are obtained from the physical solution of the system
\begin{align}
    k\bigg( \frac{a}{a+b} \bigg) &= k_m \bigg( \frac{\alpha}{\alpha + \beta} \bigg); \quad \frac{k}{\mu_p} - b = \frac{k_m}{\mu_p}; \notag  \\
    a + b &= \frac{\alpha + \beta}{\mu_p}.
\end{align}
The first two equations ensure that the mean and variance are exact in the burst limit. The third equation imposes a constraint on the total ``sharpness" of the beta distribution, thus preserving the timescale of the promoter dynamics. In Appendix B3, we show that $G_f(z)$ reduces exactly to the previously derived result in Ref. \cite{Shahrezaei-PNAS-2008} when $\frac{\mu_m}{\mu_p} \rightarrow \infty$. As a result of the fundamentally different nature of the fast mRNA limit, we note that $G_f(z)$ only yields the true Fano factor and recovers the limiting cases of fast and slow promoter dynamics if the fast mRNA limit is already true. Unlike $G_s(z)$, we obtained $G_f(z)$ in a much more heuristic manner by extrapolating from a limiting case constraint rather than taking explicit account of the trajectories of promoter history.

\section{Simulation Results}
\begin{figure*}
    \centering
    \includegraphics[width=1\linewidth]{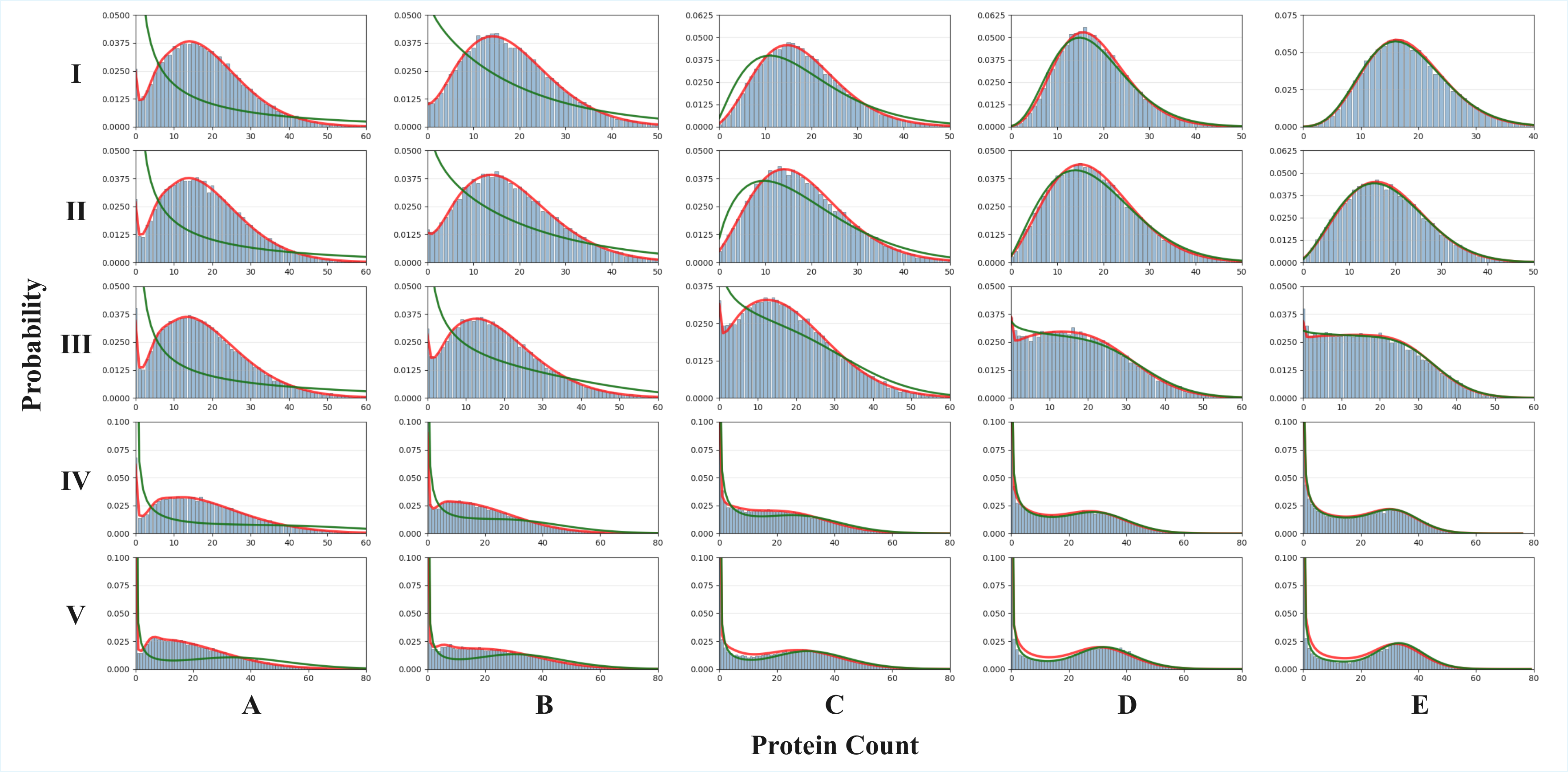}
    \caption{Histograms of simulation results plotted with the analytical approximations of the three-stage model as $(\frac{\alpha}{\mu_p};\frac{\beta}{\mu_p})$ and $\frac{\mu_m}{\mu_p}$ are varied. The \textcolor{red}{red} and \textcolor{OliveGreen}{green} curves correspond to $\color{red}{G_s(z)}$ and $\color{OliveGreen}{G_f(z)}$ respectively. $\frac{k_p}{\mu_p}=\frac{k_m}{\mu_m}=6$ are fixed to ensure a uniform mean protein count across all 25 sets of simulations. Plots in each column (labeled A-E) all have the same mRNA timescale and plots in the same row (labeled I-V) all have the same promoter timescale. The parameter ratios which set the relative timescales are 
    A) $\frac{\mu_m}{\mu_p}=\frac{1}{8}$.
    B) $\frac{\mu_m}{\mu_p}=\frac{1}{3}$.
    C) $\frac{\mu_m}{\mu_p}=1$.
    D) $\frac{\mu_m}{\mu_p}=3$.
    E) $\frac{\mu_m}{\mu_p}=8$.
    I) $\frac{\alpha}{\mu_p}=\frac{\beta}{\mu_p}=8$.
    II) $\frac{\alpha}{\mu_p}=\frac{\beta}{\mu_p}=3$.
    III) $\frac{\alpha}{\mu_p}=\frac{\beta}{\mu_p}=1$.
    IV) $\frac{\alpha}{\mu_p}=\frac{\beta}{\mu_p}=\frac{1}{3}$.
    V) $\frac{\alpha}{\mu_p}=\frac{\beta}{\mu_p}=\frac{1}{8}$.}
\end{figure*}
\begin{table*}[]
    \centering
    \begin{tabular}{|c|c|c|c|c|c|}
        \hline
            & \textbf{A} & \textbf{B} & \textbf{C} & \textbf{D} & \textbf{E}\\
        \hline
        \textbf{I}   & 0.0009242, 0.7240413 & 0.0007493, 0.2814404 & 0.0008280, 0.0579320 & 0.0008366, 0.0057970 & 0.0009318, 0.0006196 \\ 
        \hline
        \textbf{II}  & 0.0011970, 0.7485360 & 0.0012399, 0.2685857 & 0.0013037, 0.0459064 & 0.0012353, 0.0032784 & 0.0013795, 0.0013929 \\ 
        \hline
        \textbf{III} & 0.0014921, 0.6461176 & 0.0011957, 0.2386847 & 0.0014552, 0.0395966 & 0.0016529, 0.0038672 & 0.0020113, 0.0026681 \\ 
        \hline
        \textbf{IV}  & 0.0013598, 0.5826269 & 0.0020797, 0.1636233 & 0.0022986, 0.0213808 & 0.0040027, 0.0061045 & 0.0065435, 0.0053412 \\ 
        \hline
        \textbf{V}   & 0.0022045, 0.3761033 & 0.0034315, 0.0957915 & 0.0116509, 0.0126586 & 0.0164114, 0.0064237 & 0.0271033, 0.0059094 \\ 
        \hline
    \end{tabular}
    \caption{The corresponding KL divergences between the analytical approximations and simulation results from Figure 3. In each box, the first value is from using $\color{red}{G_s(z)}$ and the second from $\color{OliveGreen}{G_f(z)}$.}
\end{table*}

Given that the generating functions derived are approximate, it is critical that we make ample comparisons against numerical simulations for a broad range of parameters. Specifically, we carried out stochastic simulations for the three-stage model using Gillespie's algorithm \cite{Gillespie-JCompPhys-1976} for a fixed mean steady-state protein level (achieved by adjusting $k_m$ as necessary) and average protein lifetime $\mu_p^{-1}$ while simultaneously varying the relative timescale of the promoter ($\frac{\alpha}{\mu_p}; \frac{\beta}{\mu_p}$) and mRNA ($\frac{\mu_m}{\mu_p}$) dynamics. We quantify deviations between our analytical protein distributions $P_{\text{anal}}(n)$ and their simulated counterparts $P_{\text{sim}}(n)$ by the KL divergence  
\begin{equation}
    \mathcal{H}\big(P_{\text{sim}} \big\vert \big\vert P_{\text{anal}}\big)  = \sum_n P_{\text{sim}}(n) \ln \bigg( \frac{P_{\text{sim}}(n)}{P_{\text{anal}}(n)}\bigg)
\end{equation}
where $P_{\text{anal}}(n)$ is obtained from its generating function counterpart $G(z)$ by $\frac{1}{n!} [ \partial^nG(z)/\partial z^n ]\vert_{z=0}$. The results are shown in Figure 3 and the corresponding KL divergences are reported in Table I. 

Remarkably, $G_s(z)$ not only shows excellent agreement with the simulation data for the regions of the parameter space we expect based on the limiting cases it is exact in, but also in the intermediate and faster mRNA timescales with minimal KL divergence (Table I). $G_f(z)$ on the other hand only shows excellent accuracy for the faster mRNA timescales considered and breaks down as $\frac{\mu_m}{\mu_p}$ is decreased, which is expected. Surprisingly, even in the fastest mRNA regime $\frac{\mu_m}{\mu_p}=8$ in this initial set of simulations, $G_s(z)$ shows roughly the same accuracy as $G_f(z)$ in most cases. $G_f(z)$ only has a notable advantage over $G_s(z)$ when we simultaneously have faster mRNA dynamics and slower promoter switching. Taken together, $G_s(z)$ is a much stronger and versatile analytic approximation for the steady-state protein distribution of the three-stage model than $G_f(z)$. In subsequent simulations where we continue the trend in row III, we verified that $G_s(z)$ remains accurate for $\frac{\mu_m}{\mu_p} > 10$ and only breaks down past $\frac{\mu_m}{\mu_p}=50$. When $\frac{\mu_m}{\mu_p} \geq 10$, the distribution derived in Ref. \cite{Shahrezaei-PNAS-2008} by directly using the translational burst approximation has already been shown to be numerically accurate, making $G_f(z)$ mostly redundant.

\section{Discussion}
Although the three-stage model has been exactly solved in all limiting timescales of promoter and mRNA fluctuations, the exact protein distribution of the three-stage model away from these limiting regimes remains a notoriously difficult open problem and evades an exact solution. Recent work carried out in Ref. \cite{Wang-JChemPhys-2023} has made some notable progress where the authors solved the master equations for the time-dependent, mRNA-protein joint distribution of the three-stage model in the absence of protein degradation ($\mu_p=0$). However, for a nonzero $\mu_p$, the authors note that the master equations become analytically intractable. In this paper, using insights from the time-inhomogeneous extension of the PPA mapping, we have formulated two analytical approximations for the three-stage model as beta-mixtures of the two-stage model solution that are asymptotically exact in different limits. $G_s(z)$ is the first analytical expression that is accurate in a significant region of the parameter space absent of any small parameter ratios. In contrast, the other approximation $G_f(z)$ only shows the same level of excellent accuracy in regimes that are already well approximated by the translational burst limit \cite{Shahrezaei-PNAS-2008}. For the broad range of parameters where $G_s(z)$ is very accurate, $G_s(z)$ may also serve as a useful tool for efficient parameter inference based on steady-state protein measurements, which are more stable than mRNA measurements and are better phenotypic indicators \cite{Laurent-Prot-2010}. 

The time-inhomogeneous PPA mapping approach also gives insight into the structural difficulty of solving the three-stage model. While all realizations of the promoter history can be averaged over via a single scalar to elegantly obtain the exact mRNA distribution, this is analytically impossible for the protein distribution. At the level of the protein dynamics, each realization of the promoter history maps onto a function of $z$ instead of a scalar and although we can represent the generating function exactly as a path integral, there is no exact analytic solution for the distribution $p(\tilde{\Lambda}_z)$ which we can carry out the averaging with. Protein fluctuations are not only affected by promoter switching, but are also generated by a hidden layer of stochastic mRNA fluctuations that are subsequently integrated over the protein lifetime; our ansatze effectively approximate how hidden mRNA fluctuations induced by promoter switching are propagated to the protein level. 
In the fast mRNA limit, the hidden layer of mRNA fluctuations is eliminated and re-encoded as fluctuations in the protein burst sizes. Consequently, a separate ansatz $G_f(z)$ had to be formulated to recover this limiting case. The fundamentally different mathematical structure of $G_f(z)$ from $G_s(z)$ prevents a single analytical expression that can interpolate between all asymptotic limits. 

In future work, it will be instructive to investigate if similarly formulated approximations using different functional forms for the mixing distributions can accurately approximate protein distributions for extensions of the three-stage model with more complex promoter regulatory motifs \cite{Sanchez-PNAS-2008, Zhou-SIAM-2012} and feedback auto-regulation \cite{Jay-PRE-2014, Hornos-PRE-2005}. For such models, if the exact mRNA distribution is known, then the corresponding functional form of the mixing distribution follows \cite{Jay-PRE-2009}; we hope to extend these analytic constructions towards models of gene expression that are more biologically relevant. While the exact solution for the protein distribution for the three-stage model and even a unified approximation remains elusive, we have demonstrated that highly accurate analytical characterizations are well within reach outside of limiting cases.

\section*{Acknowledgments}
We thank Thierry Platini for his insights and efforts during the early stages of the work.

\onecolumngrid
\section*{Appendix A: Stationary distribution for $\lambda$}
\setcounter{equation}{0}
\renewcommand{\theequation}{A\arabic{equation}}
In the stationary limit, Eqs. (20) and (21) become
\begin{align}
    \beta p_1(\lambda) - \alpha p_0(\lambda) + \mu_m \frac{\partial}{\partial \lambda}\big[\lambda p_0(\lambda)\big] &= 0; \\
    \alpha p_0(\lambda) - \beta p_1(\lambda) + \mu_m \frac{\partial}{\partial \lambda}\big[(\lambda-1) p_1(\lambda)\big] &= 0
\end{align}
respectively. Combining Eqs. (A1) and (A2) gives
\begin{equation}
    \lambda p_0(\lambda) + (\lambda-1)p_1(\lambda) = K
\end{equation}
where $K$ is a constant from integration. Enforcing the physical boundary conditions that $p_0(1)=0$ and $p_1(0)=0$ in the steady-state, we must have $K=0$, leading to the very useful relation $p_1(\lambda) = \lambda p(\lambda)$. Substituting $p_0(\lambda) = p(\lambda) - p_1(\lambda)$ and then using the very useful relation, we can recast Eq. (A1) as a separable equation solely in terms of $p$
\begin{equation}
    \mu_m\lambda(\lambda-1)\frac{\partial p}{\partial \lambda} = \big[ \alpha(\lambda-1) + \beta\lambda -\mu_m(\lambda-1) - \mu_m\lambda \big] p(\lambda)
\end{equation}
with the corresponding solution $p(\lambda) \propto (1-\lambda)^{\frac{\beta}{\mu_m}-1}\lambda^{\frac{\alpha}{\mu_m}-1}$ claimed in the main text. 

\section*{Appendix B: Three-stage Model Approximations}
\setcounter{equation}{0}
\renewcommand{\theequation}{B\arabic{equation}}
\subsection*{1. Fano Factor of $G_s(z)$}
The Fano factor can be extracted from a probability generating function as
\begin{equation}
    \text{FF} = \frac{G''(1)}{G'(1)} - G'(1) + 1. 
\end{equation}
Applying the identity $\frac{d}{dx}{_1F_1} \big[a; b; cx \big] = \frac{ac}{b}{_1F_1} \big[a+1; b+1; cx \big]$, it is easy to see that
\begin{align}
    G'(1) &= \int_0^1 d\lambda\, B\bigg(\lambda; \frac{\alpha}{C}, \frac{\beta}{C}\bigg) \bigg[ \frac{k_mk_p}{\mu_m\mu_p}\lambda \bigg] \\
    G''(1) &= \int_0^1 d\lambda\, B\bigg(\lambda; \frac{\alpha}{C}, \frac{\beta}{C}\bigg) \bigg[ \bigg( \frac{k_mk_p}{\mu_m\mu_p} \lambda \bigg)^2 + \frac{k_mk_p^2}{\mu_m\mu_p(\mu_m+\mu_p)} \lambda \bigg]. 
\end{align}
Making use of the first two moments of the beta distribution and then plugging everything into Eq. (B1), 
\begin{equation}
    \text{FF}_{\text{ansatz}} = 1 + \frac{k_p}{\mu_m+\mu_p} + \frac{k_mk_p}{\mu_m\mu_p} \bigg( \frac{\beta}{\alpha+\beta} \bigg)\bigg( \frac{C}{\alpha+\beta+C} \bigg).
\end{equation}
The exact Fano factor of the protein distribution of the three-stage model is given by \cite{Raser-Science-2004}
\begin{equation}
    \text{FF} = 1 + \frac{k_p}{\mu_m+\mu_p} + \frac{k_m k_p\beta(\alpha+\beta+\mu_m+\mu_p)}{(\alpha+\beta)(\mu_m+\mu_p)(\alpha+\beta+\mu_m)(\alpha+\beta+\mu_p)}
\end{equation}
so equating FF$_{\text{ansatz}} =$ FF, we find that 
\begin{equation}
    \frac{1}{C} = \frac{1}{\mu_m} + \frac{1}{\mu_p} - \frac{1}{\alpha + \beta + \mu_m + \mu_p}.
\end{equation}

\subsection*{2. Limiting Cases of $G_s(z)$}
(1) In the limit of fast promoter switching where $\alpha \rightarrow \infty$ and $\beta \rightarrow \infty$ while keeping $\frac{\alpha}{\alpha+\beta}$ constant, the beta distribution becomes a delta function centered at $\frac{\alpha}{\alpha+\beta}$. Integrating over $\delta(\lambda - \frac{\alpha}{\alpha+\beta})$ in place of the beta distribution in Eq. (36) gives
\begin{align}
    G_s(z) = \lim_{N\rightarrow \infty} \exp \bigg\{ N\bigg( &{_1F_1} \bigg[\frac{k_m/N}{\mu_p} \bigg( \frac{\alpha}{\alpha+\beta}\bigg); \frac{\mu_m}{\mu_p}; \frac{k_p}{\mu_p}(z-1)  \bigg] - 1 \bigg) \bigg\}
\end{align}
which corresponds to the two-stage model \cite{Pendar-PRE-2013} with the rescaled transcription rate $k_m \rightarrow k_m \big( \frac{\alpha}{\alpha+\beta}\big)$.

(2) When promoter switching is slow where $\alpha \rightarrow 0$ and $\beta \rightarrow 0$ while holding $\frac{\alpha}{\alpha+\beta}$ constant, the beta distribution for small parameters becomes increasingly concentrated at its boundaries, leading to the correlated limit
\begin{equation}
    \lim_{(a, b) \rightarrow (0,0)} B \big( \lambda; a, b \big) = \bigg( \frac{b}{a+b} \bigg) \delta(\lambda) + \bigg( \frac{a}{a+b} \bigg) \delta(\lambda-1).
\end{equation}
Using this identity in Eq. (36) and integrating over it, we arrive at
\begin{equation}
    G_s(z) = \frac{\beta}{\alpha+\beta} + \bigg(\frac{\alpha}{\alpha+\beta}\bigg) G_{\text{two-stage}}(z),
\end{equation}
which is a linear combination between the generating functions of the adiabatic promoter states weighted by their steady-state fractional occupancy. 

(3) In the limit of slow mRNA, the protein dynamics reach a fast equilibrium \cite{Swain-JMathBiol-2016}. Here, the protein distribution can be obtained from the mRNA distribution as 
\begin{align}
    G(z) &= G_m \bigg(\exp\bigg\{\frac{k_p}{\mu_p}(z-1)\bigg\}\bigg) \notag \\
    &= {_1F_1} \bigg[ \frac{\alpha}{\mu_m}; \frac{\alpha+\beta}{\mu_m}; \frac{k_m}{\mu_m}\bigg(\exp\bigg\{\frac{k_p}{\mu_p}(z-1)\bigg\}-1\bigg) \bigg].
\end{align}
We proceed to show that Eq. (36) reduces to Eq. (B10) for $\frac{\mu_m}{\mu_p} \rightarrow 0$. Using the identity for small parameters of ${_1F_1}$ in the correlated limit \cite{Slater-confluent}
\begin{equation}
    \lim_{(a,b) \rightarrow (0,0)} {_1F_1}[a;b;z] = 1 + \frac{a}{b}(e^z-1)
\end{equation}
in Eq. (36) yields
\begin{align}
    G_s(z) = \int_0^1 d\lambda \, B\bigg(\lambda; \frac{\alpha}{C}, \frac{\beta}{C}\bigg) \exp \bigg[ \frac{k_m}{\mu_m} \lambda \bigg(\exp\bigg\{ \frac{k_p}{\mu_p}(z-1) \bigg\} -1 \bigg) \bigg].
\end{align}
For small $\mu_m$, $1/C$ scales according to $\mathcal{O}(1/\mu_m)$, thus $C \sim \mu_m$. From the identity that the generating functions in Eqs. (23) and (24) are equivalent \cite{Jay-PRE-2009}, the equivalence between Eqs. (B10) and (B12) follows. 

\subsection*{3. Fast mRNA limit of $G_f(z)$}
Let us define the dimensionless quantity $\theta = \frac{\mu_m}{\mu_p}$. Enforcing the burst limit $\theta \rightarrow \infty$ in the ${_1F_1}$ function part of Eq. (38), it reduces to \cite{Slater-confluent}
\begin{equation}
    \lim_{\theta \rightarrow \infty }{_1F_1}\bigg[ \frac{k/N}{\mu_p}; \theta; \frac{k_p }{\mu_m}\theta\lambda (z-1) \bigg] = \bigg[ \frac{1}{1-\frac{k_p}{\mu_m}\lambda(z-1)} \bigg]^{\frac{k/N}{\mu_p}}
\end{equation}
which notably corresponds to the generating function of a negative binomial distribution. Now, we have
\begin{align}
    G&_f(z) = \lim_{N\rightarrow \infty} \int_0^1 d\lambda \, B\big(\lambda; a, b \big)  \exp \bigg\{ N \bigg(\bigg[ \frac{1}{1-\frac{k_p}{\mu_m}\lambda(z-1)} \bigg]^{\frac{k/N}{\mu_p}} - 1 \bigg) \bigg\}.
\end{align}
Rewriting the exponential function part of Eq. (B14) in terms of the identity $\exp\{g(z)\} = \big( 1 + \frac{g(z)}{N} \big)^N$ as $N \rightarrow \infty$
\begin{align}
    G&_f(z) = \int_0^1 d\lambda \, B\big(\lambda; a, b \big) \bigg[ {1-\frac{k_p}{\mu_m}\lambda(z-1)} \bigg]^{-\frac{k}{\mu_p}}
\end{align}
which we recognize as the Euler integral representation of the Gaussian hypergeometric function \cite{NIST}. Mapping the relevant parameter combinations to the arguments of ${_2F_1}$ function and using the physical solution of $\{a,b,k\}$ from Eq. (39), we recover the result derived in Ref. \cite{Shahrezaei-PNAS-2008} as claimed in the main text.

\twocolumngrid

\end{document}